# *i*Phantom: a framework for automated creation of individualized computational phantoms and its application to CT organ dosimetry

Wanyi Fu, Shobhit Sharma, Ehsan Abadi, Alexandros-Stavros Iliopoulos, Qi Wang,

Joseph Y. Lo, Xiaobai Sun, William P. Segars, Ehsan Samei

*Abstract*— *Objective: This study aims to develop and validate a novel framework, iPhantom, for automated creation of patient-specific phantoms or "digital-twins (DT)" using patient medical images. The framework is applied to assess radiation dose to radiosensitive organs in CT imaging of individual patients. Method: From patient CT images, iPhantom segments selected anchor organs (e.g. liver, bones, pancreas) using a learning-based model developed for multi-organ CT segmentation. Organs challenging to segment (e.g. intestines) are incorporated from a matched phantom template, using a diffeomorphic registration model developed for multi-organ phantom-voxels. The resulting full-patient phantoms are used to assess organ doses during routine CT exams. Result: iPhantom was validated on both the XCAT (n=50) and an independent clinical (n=10) dataset with similar accuracy. iPhantom precisely predicted all organ locations with good accuracy of Dice Similarity Coefficients (DSC) >0.6 for anchor organs and DSC of 0.3-0.9 for all other organs. iPhantom showed <10% dose errors for the majority of organs, which was notably superior to the state-of-the-art baseline method (20-35% dose errors). Conclusion: iPhantom enables automated and accurate creation of patient-specific phantoms and, for the first time,*

*provides sufficient and automated patient-specific dose estimates for CT dosimetry. Significance: The new framework brings the creation and application of CHPs to the level of individual CHPs through automation, achieving a wider and precise organ localization, paving the way for clinical monitoring, and personalized optimization, and large-scale research.*

## I. INTRODUCTION

COMPUTATIONAL human phantoms (CHPs) are mathematical representations of the human anatomy in a digital format. CHPs and their applications have co-evolved in the last six decades [1, 2]. CHP development is driven by important and growing applications, which include retrospective, prospective, or real-time radiation dosimetry, individual cancer risk estimation, diagnostic and interventional radiology studies, monitoring for environmental radiation exposure, assessment of medical imaging protocols, design and verification of shielding protection, and virtual clinical trials for regulatory submissions[1-10]. For these applications, it is essential to have CHPs that realistically reflect individual patients (i.e., patient-specific) as well as the population at large (i.e., population-specific) to echo anatomical variations of real clinical cases and scenarios.

Toward that aim, recent development of CHPs has focused on realistically representing given individuals for patient-specific investigations or assembling many anatomically variable phantoms at large scale for population-based studies. The phantoms are primarily developed by manually segmenting a limited number of patient cases. However, segmentation is a time-consuming process that can take many months to complete per phantom depending on the level of detail required. Furthermore, the work involves inter- and intra-operator variability that incorporates a certain degree of subjectivity in the process. Even addressing these limitations, the current automatic segmentation techniques based on tissue texture patterns and/or manually-labeled training resources are still limited to only a handful of organs with high-contrast [7, 10-12]. They are generally incapable of differentiating small organs or adjacent organs with similar textures or gray scale. Thus, segmentation alone is inadequate to reflect the extensive

This work was supported in part by the Research Grant through the National Institutes of Health under Grant R01EB001838.

W. Fu and E. Abadi are with the Department of Electrical and Computer Engineering, and Carl E. Ravin Advanced Imaging Laboratories, Duke University, Durham, NC, 27705 USA (email: wanyi.fu@duke.edu; ehsan.abadi@duke.edu).

S. Sharma is with the Department of Physics and Carl E. Ravin Advanced Imaging Laboratories, Duke University, Durham, NC, 27705 USA (email: shobhit.sharma@duke.edu).

A. Illiopoulos and X. Sun are with the Department of Computer Science, Duke University, Durham, NC, 27708, USA (email: ailiop@cs.duke.edu; xiaobai@cs.duke.edu).

Q. Wang with the Department of Radiology, the Fourth Clinical Hospital of Hebei Medical University, Heibei, 050011, China (email: wq20@hotmail.com)

J. Y. Lo is with the Departments of Electrical and Computer Engineering, Biomedical Engineering, Medical Physics Graduate Program, and Carl E. Ravin Advanced Imaging Laboratories, Duke University, Durham, NC, 27705 USA (email: joseph.lo@duke.edu).

W. P. Segars is with the Departments of Biomedical Engineering, Medical Physics Graduate Program and Radiology and the Carl E. Ravin Advanced Imaging Laboratories, Duke University, Durham, NC, 27705 USA (e-mail: paul.segars@duke.edu).

E. Samei is with the Carl E. Ravin Advanced Imaging Laboratories, the Medical Physics Graduate Program, the Departments of Radiology, Electrical and Computer Engineering, Biomedical Engineering, and Physics, Duke University, Durham, NC, 27705 USA (email: ehsan.samei@duke.edu).

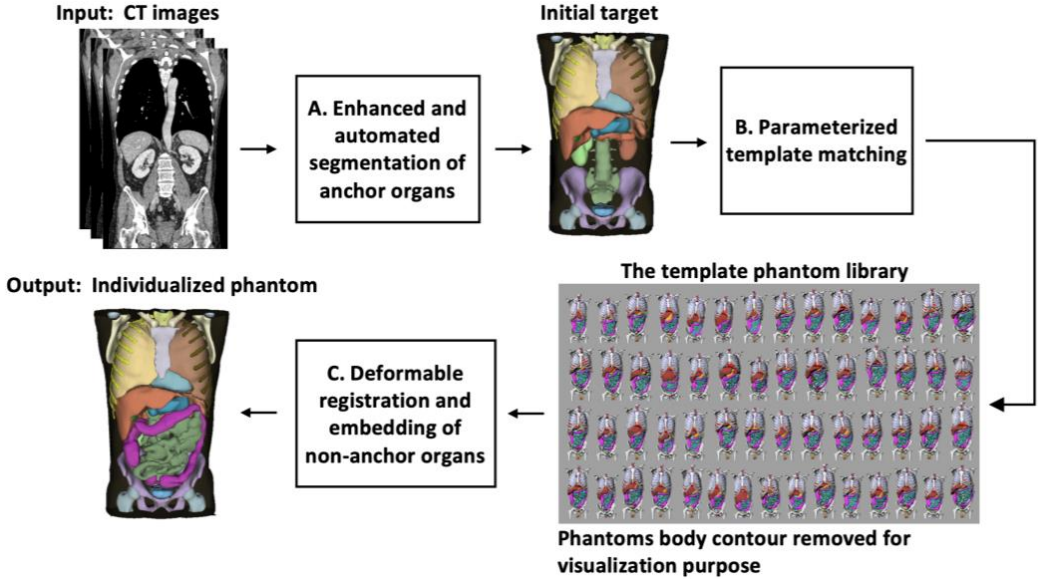

**Input: CT images** → **A. Enhanced and automated segmentation of anchor organs** → **Initial target** → **B. Parameterized template matching** → **The template phantom library**

**Output: Individualized phantom** ← **C. Deformable registration and embedding of non-anchor organs** ←

Phantoms body contour removed for visualization purpose range of organs essential for representing the human body. This necessitates a combination of segmentation and deformation to create CHPs.

Fig. 1. Flow-chart of the *i*Phantom framework.

To facilitate the development of CHPs, deformable techniques have been used to augment existing phantoms to model additional anatomically variable models [1, 2]. Populations of new models can be created by deforming given template phantoms to match variations observed in patient data. However, these methods are still highly constrained by the segmentation process. Overall, the current methods cannot be applied to create large sets of CHPs or to create digital representations, so-called digital twins, of clinical cases.

In this study, we develop a framework, *i*Phantom, to address the challenging problem of automatically and accurately creating CHPs in a patient-specific manner toward large scale CHP development and clinical patient-specific implementation. The work draws upon a validated non-automatic pipeline that developed the widely used XCAT phantoms [13-15]. Specifically, the work addresses two major obstacles to generate phantoms directly from patient medical images. The first obstacle is to extract patient specific information. We developed a machine learning model segmenting the key, so-called anchor, organs automatically from medical images. Second, a complete CHP requires all major organs, including the low contrast ones. We adapted a deformation morphing technique developed for the multi-organ registration problem in phantom space to embed unsegmented structures from an anatomical template. The template was obtained using a parameterized matching method to draw upon the complete human models from the XCAT phantom library. We systematically validate the framework in terms of phantom geometry and its application for organ radiation dosimetry. This specific application is motivated by the strong need to assess radiation dose and its associated risk to the patient during an exam in a patient-specific manner. Organ dose has been regarded as the most relevant metric to characterize patient risk. However, clinical quantification of this metric has been hindered due to the limitation in creating accurate CHPs.

In this paper we detail a process to create CHPs from patient images automatically. The preliminary concept was first introduced at the 2018 SPIE Medical Imaging conference [16]. The present manuscript significantly extends that concept into a framework with each individual component specifically developed for the purpose of creating patient-specific CHPs. The framework is further applied to the task of patient-specific organ dose estimation by incorporating a recently released GPU-based Monte Carlo simulation package for CT scans with both fixed and modulated tube current. We demonstrate the clinical utility of the framework by applying it to unseen clinical patient images to create a new set of patient-specific phantoms and quantitatively validate the results.

The rest of the paper is organized as follows. We introduce the *i*Phantom pipeline and analysis in Section II. Then, we describe its application to radiation dosimetry in CT in Sections III. The validation experiments and results are reported in Sections IV and V. Finally, we present conclusions and discussions in Section VI.

## II. *i*PHANTOM FRAMEWORK

The proposed framework aims at creating patient-specific phantoms directly from patient-medical images in a fully automated pipeline. The framework first automatically segments a selected set of organs and structures. These organs are those rendered in high contrast in CT, which can be delineated or segmented with high fidelity. The segmentation of the remaining organs, however, may suffer from a great degree of uncertainty. Thus, the framework fuses the initial segmentation results with an anatomical template that will be used to fill in the missing structures in the newly made phantoms. This component is necessary since the body consists of many organs (34 organs investigated in this study), large and small, with different morphology and textures; reliably segmenting all of them automatically (even manually) is very challenging. For this work, we used for templates the XCAT phantom library of 60 highly detailed adult models, developed in our laboratory and widely used for many applications [13, 14].

Figure 1 summarizes the steps of the *i*Phantom framework. First, anchor organs within a given set of patient CT data are automatically segmented, using a learning-based segmentation model developed to segment key organs, to define an initial target (Figure 1. A.). This person-specific definition of anchor organs is used to guide the fill-in of non-anchor organs. An XCAT phantom template that best matches the partially segmented target is then selected using parameters chosen to reflect anatomical similarities (Figure 1. B.). Finally, a mapping between the template XCAT model and the patient target was calculated using a registration model adapted for the multi-organ scenario of the phantom space. The mapping is used to transport the non-anchor organs from the template space to define them within the new patient model (Figure 1. C.). In this study, we developed both linear (affine) and nonlinear (diffeomorphic) mapping methods.

### A. Automated segmentation of anchor organs

Accurate anchor-organ segmentation is critical to the automated creation of individualized phantoms, setting the stage for the subsequent steps. The specific set of anchor organs were determined based on state-of-the-art multi-organ segmentation of CT images of the chest-abdominal-pelvis region [11, 12] and our available labeled training images. Twenty-two organs and structures were selected: thyroid, lung (L/R), heart, liver, spleen, kidney (L/R), gallbladder, ribs (L/R), bladder, spine, clavicles, sternum, scapular, stomach, pancreas, pelvis, femur, arm, and body. The body represents all organs and tissues not individually segmented but included within the body contour.

3D convolutional neural networks were developed using a Unet architecture is similar to that described by Çiçek et al. [17] and detailed in the Appendix Section I.A. In designing the training objective, or the loss function, we considered multiple classes (organs) as well as the difference in organ volume, i.e., the number of voxels in an organ. For example, the ratio in volume between the lung and thyroid can be up to three orders of magnitude. To overcome the inter-class imbalance problem, we made use of the combined dice loss and cross-entropy loss functions similar to Taghanaki et al. [18]. The dice loss has been used for multi-organ segmentation, even though it may lead to failure to converge for small organs due to the vanishing gradient. The cross-entropy loss regularizes the objective function [18]. Specifically, we used the following loss function

$$Loss = w_1 \left(1 - \frac{1}{L}\sum_l \left(\frac{2\sum_i p_l^i r_l^i}{\sum_i p_l^i + r_l^i}\right)\right) - w_2 \sum_l r_l^i \log(p_l^i r_l^i), \quad (1)$$

where $w_1$ and $w_2$ are weighting coefficients for the dice and cross entropy loss, respectively; $p_l^i$ and $r_l^i$ are the segmentation probability and binary indicator, respectively, for voxel $i$ and class $l$. In this study, we chose $w_1 = 1$ and $w_2 = \frac{1}{L}$, where L is the total number of classes.

For training the network, we utilized and refined the manually segmented CT data upon which the XCAT library of phantoms were based. In practice, the GPUs we used had limited memory resources. To account for this, we cropped the training images to 128x128x128. Within this input size constraint, to balance sufficient global content for training and resolution for creating phantoms, the images were down-sampled to a longitudinal resolution of 5 mm and an in-plane resolution of 2.5 mm. The inputs were at the size that contained the majority volume of the patient trunk. In the training, CT images were randomly sampled with each structure centered, following Pawlowski et al., to ensure all the structures were trained [12]. For inference, the whole CT images, rather than cropped segments, were used as inputs to speed up prediction and eliminate prediction window boundary artifact. Once developed, the segmentation method was validated as described in Section IV.A.

### B. Parameterized template matching

After segmentation, a parameterized matching strategy was developed to identify a template phantom that best matches the patient determined by the segmented anchor anatomy. It was assumed that if the anchor layout is similar between a patient and a phantom, the rest of the organs will show a higher likelihood of similarity. This similarity also results in the transformation computation being less expensive. Thus, we find phantom i that minimizes the distance to the patient defined by parameters $\Theta_i = \{\theta_{1,i}, \theta_{2,i}, ... \theta_{k,i}\}$ derived from the anchor organs as

$$i = arg \min_i \|\Theta_i - \Theta_0\|_2,$$
$$\text{subject to } i \in G, \quad (2)$$

where $\Theta_i$ and $\Theta_0$ are anatomical parameters for phantom i and the target patient, respectively, and $G$ is the set of phantoms satisfying a constraint (e.g., age range, gender).

In previous work, the trunk height has been shown to be a good indicator of organ distribution [19]. Likewise, the trunk diameter has been shown to be related to the thickness of tissue outside the skeleton [20]. Therefore, in this initial implementation, we defined $\theta_{1,i}$ as the phantom trunk height, and $\theta_{2,i}$ as the phantom trunk effective diameter defined as $\theta_2 = 2\sqrt{\frac{V}{h\pi}}$, where V and $h$ are the segmented trunk contour volume and height, respectively. The matched phantom was then chosen iteratively from $G$ using (2).

### C. Registration and embedding of non-anchor organs

With a matching template phantom, the final step was to calculate a mapping from the template space to the target patient space. A mapping may be described as the product composition of an affine mapping for initial global alignment and a non-rigid, non-linear diffeomorphic mapping. The affine mapping has only shifting, scaling and sheering parameters and can be determined fast. The affine-transformed template is then used for the diffeomorphic registration to the target. The diffeomorphic transformation enables large deformation calculations while preserving the topology. We developed and evaluated mapping calculations with both 1) affine alone and 2) combined affine with diffeomorphic mapping (labeled as diffeomorphic mapping).

From different types of deformable registration models and methods that exist today [21], we adapted the Advanced Normalization Tools (ANTs), underlain by the symmetric normalization (SyN) method, recognized for its reliable accuracy and wide use for medical research [22, 23]. We modified the transformation calculation parameters from its common utility in brain image registration to calculate the mapping from the template to the target. The mapping was applied to the non-anchor organs located in the template to 'fill-in' these organs in the target.

The registration calculation adapted the SyN method and we adjusted parameters including similarity metrics and their associated parameters, a gradient step, and Gaussian smoothing for velocity and deformation field. The parameters were searched by conducting a large set of experiments over possible parameters to identify those with the sufficient registration accuracy. The similarity metrics between the phantoms were calculated in both intensity-based and label-based spaces. Intensity-based metrics are advantageous for regions with more drastic changes (e.g. soft tissue and bone boundaries) while label-based metrics provide additional anatomical similarity knowledge. For intensity-based phantom input, we generated synthesized CT images from the template and target voxel phantoms by assigning organ intensities (HU values) derived from the averaged organ voxels of the XCAT patient CT images. For label-based phantom input, the template and target voxel phantoms were formatted with corresponding unique IDs assigned to the segmented structures. For organs with two separate regions on both left and right sides of the body (e.g., lungs), unique labels were assigned for each side. For the affine transform, we used the intensity-based mutual information metric. For the deformable transform, we used the combined intensity-based cross-correlation and label-based point-set expectation metrics. The two metrics were weighted and summed (0.1/0.9). For cross-correlation, we specified a neighborhood of 27 voxels for an efficient sliding window-based local cross-correlation calculation. The point set expectation calculates the weighted sum of distances of a voxel in one image to a set of voxels in the other image [24]. The weighting was determined by the distance in a normal function within a neighborhood of 20 voxels.

## III. THE APPLICATION TO PATIENT-SPECIFIC ORGAN DOSIMETRY

The phantoms created using the framework are portable to applications in many domains. In this study, we applied and validated the framework in the classic and actively researched area of *patient-specific* organ dosimetry in computed tomography. Computed tomography has been widely used for diagnosis of major diseases; however, its potential harmful radiation effect has been a concern for healthcare providers and patients. Patient-specific organ dose has been regarded as the most relevant metric to quantify radiation exposure and the associated risk. However, it has not been widely utilized clinically due to the challenge of automatically creating patient-specific phantoms necessary to estimate credible organ doses. We attempted to address this limitation.

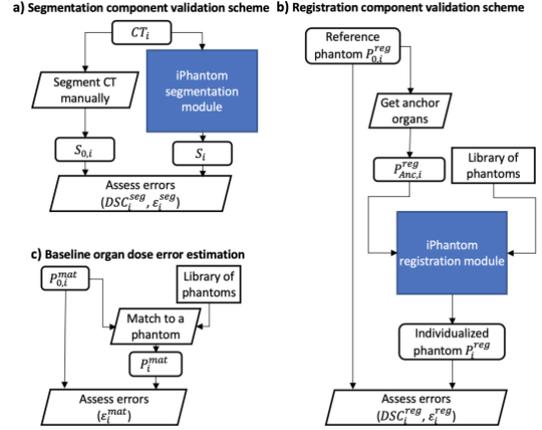

Fig. 2. Validation scheme for the XCAT experiments with a) the segmentation component, b) registration component, and c) the baseline organ dose error estimation. Oval represents data, parallelogram represents operation, rectangular represents pre-computed functions. Superscripts seg, reg, mat represent studies for segmentation, registration, and matching respectively. Subscripts 0 represents reference, and i represents the $i^{th}$ phantom. The symbols are listed in Table 1.

TABLE 1. TABLE OF SYMBOLS

| | |
|---|---|
| $i$ | The $i^{th}$ phantom generated by the test set |
| $0, i$ | The $i^{th}$ reference phantom |
| anc | Phantom with anchor organs only |
| non-anc | Phantom with non-anchor organs |
| MAE | Mean absolute relative error across all test set |
| DSC | Dice similarity coefficients |
| $DSC_i^{seg}, DSC_i^{reg}$ | DSC between the $i^{th}$ test set and reference phantom for segmentation and registration validation, respectively |
| $\varepsilon_i^{seg}, \varepsilon_i^{reg}, \varepsilon_i^{mat}$ | MAE between the $i^{th}$ test set and reference phantom for segmentation validation, registration validation, and matching method, respectively |

For this validation, we utilized a CT organ dose estimation module [25] developed in our laboratory. As input, the patient-specific phantoms were converted into voxelized dosimetry phantoms by assigning materials to each organ and structure. The CT technical parameters are specified in Section IV D, including vendor and CT geometry, bowtie, spectrum, and tube current profile. The absorbed dose was obtained using a validated real-time Monte Carlo (MC) tool, developed using the MC-GPU framework [26], to simulate photon transport across the voxelized dosimetry phantoms.

In general, organ dose estimation requires different levels of accuracy in terms of the phantom geometry for fixed and modulated tube current. It has been demonstrated that the organ dose estimation under fixed tube current is reasonably accurate as long as the patient is matched to a phantom with similar size, as the radiation field under fixed tube current is relatively uniform [20]. Organ dose estimation under the more prevalently used TCM requires the organs localized more closely to the patient, as the radiation field is more heterogeneous. We evaluated organ dose estimation accuracies under both fixed and modulated tube current.

## IV. Validation

We systematically validated the *i*Phantom framework using the XCAT phantoms and the CT data upon which they were based (XCAT datasets) in a cross-validation approach. We further applied the framework to clinical CT images to evaluate the framework's generalizability in creating new patient-specific phantoms.

For the XCAT datasets, we validated the segmentation stage and data fusion stage (matching and registration) individually. The XCAT datasets were not directly used to validate the full pipeline because the XCAT geometries have been generalized and altered to adapt to many applications (e.g., respiratory and cardiac motion simulations) and thus are no longer fully aligned to their original CT images. We took advantage of the XCAT datasets consisting of an extensive range of organs within a population to validate individual components of the *i*Phantom framework. We used the manually segmented CT data upon which the XCAT library of phantoms were based to validate the anchor organ segmentation accuracy. We further adopted the voxelized XCAT phantoms to assess the registration calculation and non-anchor organ embedding of the framework. A schematic of the validation strategy for the segmentation validation, registration validation, and a baseline dose error estimation is shown in Figure 2.

### A. Segmentation validation

We used fifty chest-abdomen-pelvis CT datasets that were part of the XCAT modeling. We further refined and checked their original manual segmentation under supervision of a radiologist (23 years of experience). The image data show clinical anatomical variations with no abnormalities (20/30 F /M; age range: 18 - 78 y.o.; trunk effective diameter range: 24 - 39 cm). The twenty-two organs and structures listed in Section II.A were segmented in the data.

A five-fold cross-validation was performed to train and validate the segmentation models. Each testing CT dataset was automatically segmented using the trained model of each fold and compared to its corresponding expert manual segmentation in terms of geometrical accuracy and estimated dose (metrics described in Section IV.D.). In each fold, the training, validation, and test set were divided as 30, 10, and 10 of the cases, respectively. The model was implemented by Keras with a Tensorflow backend with Adam optimizer. For training, we used 24,000 iterations with a learning rate of 10e-3 and 10e-4 for the first and second half of iterations, respectively. Training took about 20 hours, and the prediction of one patient took about 1-20 seconds using a Titan RTX GPU with 24 GB memory.

### B. Registration validation

To validate the registration accuracy, we performed two experiments: 1) a leave-one-phantom-out, and 2) a leave-one-organ-out. The leave-one-phantom-out approach was used to assess the ability of the framework in predicting the organs for unknown targets. The leave-one-organ-out approach was used to evaluate the accuracy of predictions with different anchor organs left out. The anchor organs from the XCAT phantoms were created based on segmentation, while the non-anchor organs were derived from a previously developed, not-segmentation-based registration approach. The leave-one-organ validation attempted to evaluate the accuracy with segmentation-based ground truth.

Furthermore, to evaluate the isolated effect of diffeomorphic deformation, we assessed the two registration methods, one using only an affine transformation and one with the combined affine and diffeomorphic transformation (labeled as diffeomorphic deformation in the results).

For this validation, we voxelized 50 XCAT phantoms from the existing library to obtain anchor and non-anchor organ phantom voxels (organ types specified in each experiment). To simulate the prior segmentation step, the target phantom was set to resemble the "segmented" image by setting each of the anchor organs to unique integer IDs with the rest of the structures set to a body ID. The matched XCAT was set up in the same manner to create a corresponding template image. Given these images, the framework was used to calculate the transform from the template to the target. The transformation was applied to the template with full anatomy to predict the remaining anatomy of the target, compared to the original target phantom, using the metrics as outlined in Section IV. D. The procedure for each target phantom is detailed in Fig 2b.

For computational efficiency in terms of the registration as well as the dosimetry calculations, in all registration experiments, the XCAT phantoms were voxelized at an isotropic resolution of 3.45 mm.

#### 1) Leave-one-phantom-out experiment

For this experiment, the anchor organ types were those specified in Section II A. The non-anchor organs included all other radiosensitive structures: thymus, larynx pharynx, trachea bronchi, esophagus, breasts, large intestine, adrenals, small intestine, ovaries, testes, uterus, and vagina.

In the leave-one-phantom-out experiment, each of 50 XCAT phantoms was used as a target while the remaining 49 were used as the template library. Each target XCAT was matched to a template from the remaining 49 using the methods described in II.B.

#### 2) Leave-one-organ-out experiment

Within each of the 50 leave-one-phantom-out experiments, we performed multiple leave-one-organ-out experiments. In each leave-one-organ-out experiment, one segmented organ was left out and the rest of the organs were used as the anchor organs. The left-out organ was filled in using the *i*Phantom registration module and assessed for accuracy using metrics described in Section IV. D.

### C. Application to new CT data

The *i*Phantom framework was tested using new CT datasets, not previously used in the XCAT phantom creation. This HIPPA compliant study included ten randomly selected patients who underwent Chest-abdomen-pelvis CT scans from our institution from January 2017 to May 2017. Ten chest-abdomen-pelvic CT images were included in this experiment (4F/6M; age range: 35-83 y.o.; trunk effective diameter range: 24 - 44 cm). For each patient, the anchor organs used in the 1st experiment were manually delineated by a radiologist with 23 years of experience. This annotation was used as the evaluation reference. The *i*Phantom framework was applied to these clinical images to generate patient-specific phantoms and to evaluate the framework accuracy.

Furthermore, using the CT images and the manual annotations as reference, we evaluated the framework accuracy for segmentation and registration components using the metrics described in the Section IV. D. In detail, the trained segmentation model by the XCAT dataset (Section IV. A.) was applied to these new CT images to predict organ labels. The predicted and reference segmentation were compared by both geometry and organ dose using the same scheme as in Fig. 2a. The registration model was evaluated using the leave-one-organ-out approach. We used both the manual (reference) and the predicted segmentation as the initial target to evaluate the isolated registration error and the overall segmentation and registration error, respectively. The initial target phantom was then matched to an XCAT and had the left-out organ filled in using the proposed registration module. The template phantom libraries consisted of all the 50 XCAT phantoms with all anchor organs as used in section IV.B. 1). The filled-in organs were compared with its manual delineation using the scheme shown in Fig. 2b. For organ dose estimation, we extracted the actual tube current modulation profile from the CT image DICOM headers to mimic the scans more realistically.

### D. Validation metrics

For each study, comparisons of the predicted anatomy versus the known truth anatomies were made in terms of the geometrical accuracy of the organs and structures as well as estimated CT radiation dose. Geometrical accuracy of the predicted anatomy as compared to the known truth was measured by the dice similarity coefficients (DSC).

We measured the dose differences between the phantoms generated from each test set and its reference using the organ dose module. Without loss of generalizability, the organ doses were estimated using a typical clinical protocol. This simulation included a Light Speed VCT scanner (GE Healthcare) with explicitly modeled gantry geometry, bowtie, and spectrum. A chest-abdominal-pelvis protocol was performed with the scan converging 1 cm above the lung and 1cm below the pelvis. The CT techniques were 120 kV, pitch of 1.375, and collimation of 40 mm for simulated scans with both modulated and fixed tube current. The tube current modulation profile was synthesized as a function of patient attenuation at each projection angle, scanner-specific geometry, bowtie type, and kV using the method described by Li et al [27]. In this study, we used the 'strong' TCM configuration (i.e., $\alpha = 1$ in Li et al. [27]) to simulate a scenario with the most heterogeneous radiation field providing an upper bound in organ dose discrepancy.

The organ dose differences between the dose from the $i^{th}$ reference phantom ($D_{0,i}$) and that from the phantom generated from the test set ($D_i$) were compared by absolute relative error (ARE) as

$$\varepsilon_i = \frac{|D_i - D_{0,i}|}{D_{0,i}} 100\%, \tag{3}$$

with the mean of $\varepsilon_i$ calculated across all the tests (MAE).

### E. Baseline method

We assessed organ dose errors using an alternative baseline method to provide a dose accuracy reference. In many state-of-the-art approaches in automatically estimating organ doses, the patient anatomy is represented by a matched computational phantom without further post-processing (e.g., registration) [19, 28]. Thus, for this baseline comparison, we matched each phantom (reference) to another phantom (predicted anatomy) in a leave-one-phantom-out validation approach across the 50 XCAT phantoms. The matched phantom was selected using the proposed parametrized matching method (Section II.B). Instead of using affine or diffeomorphic registration to align the phantoms to the initial target, the matched phantom was aligned to the patient using anatomical landmarks (top of the lung and bottom of pelvis). We reported organ dose differences between each reference phantom and its matched phantom. This method reasonably represented the state-of-the-art in that 1) the XCAT phantom population is relatively large within the literature, and 2) the matching includes two decisive parameters (i.e., height and width) that influence organ dose.

## V. RESULTS

In this section, we present the validation results. We first show geometry and organ dose estimation accuracy of the segmentation and data fusion stages, respectively, using the

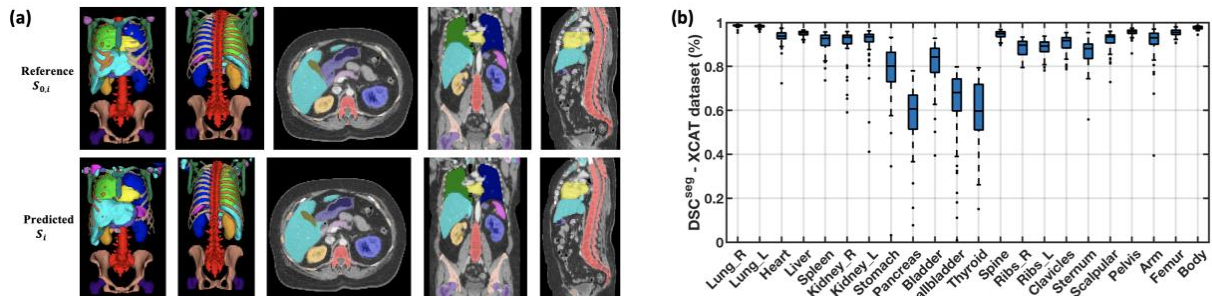

Fig. 3.   Gemometry valiation results of the segmentation component using XCAT datasets in a five-fold cross validaiton on the test set. a) Example segmentation results overlaid onto the patient CT data for patient with medican cross-organ average DSC. B) Box plot summarizing the DSC results calculated from the reference and predicted segmentations from each test case.

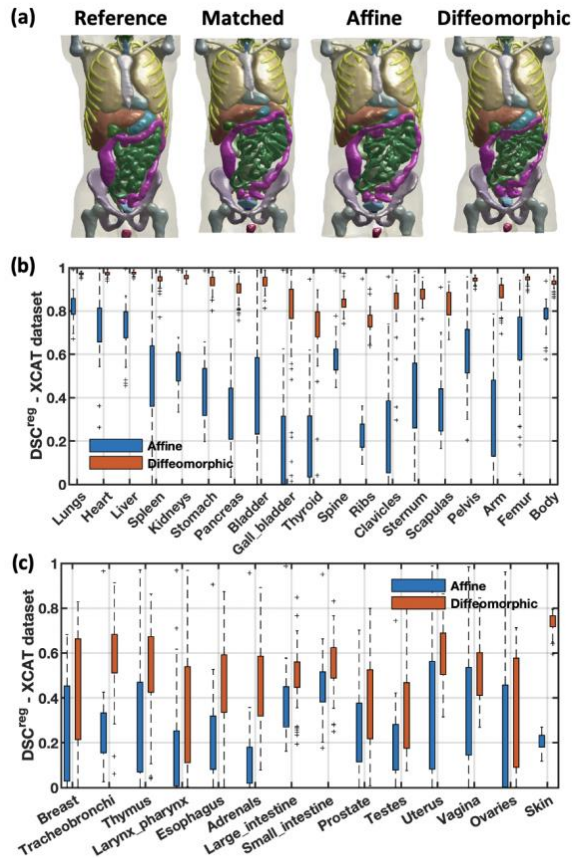

(a) Reference | Matched | Affine | Diffeomorphic

(b)

(c)

Fig. 4. Results of the registration component validation using the leave-one-phantom-out approach with the XCAT dataset. a) Rendering of a reference, the matched, and the matched model with affine and diffeomorphic transformations. b, c) Box plot of dice similarity coefficients (DSC) between the reference and matched with transformation for b) anchor organs and c) non-anchor organs for all XCAT phantoms. The results show that the diffeomorphic transformation improves the prediction for the filled in organs.

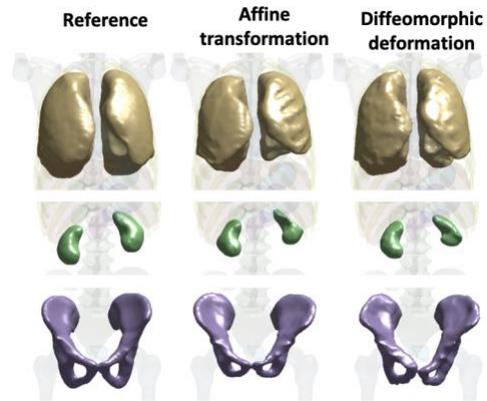

Fig. 5. Registration geometry-based validation of the XCAT dataset using the leave-one-organ-out approach. Rendering of an example case (same case as in Fig. 4) where the lungs (top), kidneys (middle), and pelvis (bottom) are left out respectively and filled in using affine transformation and diffeomorphic deformation.

XCAT dataset. We further provide overall dose error surrogates combining segmentation and registration dose errors and compare them with the baseline dose errors. Finally, we show the clinical data validation results. The supplemental tables and figures for the XCAT and clinical datasets were presented in Appendix section I.B. and I.C., respectively.

## A. Geometry validation

**Segmentation.** Fig. 3 shows the results of segmentation validation using the XCAT dataset. The results show a sufficient segmentation performance, especially in clear discrimination of organ material types as soft-tissue, bone, and lung, which are essential for accurate dose calculation. Specifically, large structures (e.g., lung and body) show an average DSC greater than 0.98. Large soft tissue organs (liver, spleen, kidneys) show an average DSC greater than 0.9. Soft tissue organs with relatively irregular shape (pancreas, bladder) or smaller size (thyroid, gallbladder) show an average DSC of 0.6-0.8. Bones show an average DSC > 0.85.

**Registration leave-one-phantom-out.** Fig. 4 shows the results of registration validation using the leave-one-phantom-out approach. For most anchor organs, affine transformation and diffeomorphic deformation resulted in a DSC of 0.2-0.6 and 0.8-0.9, respectively. For filled in organs, affine shows a reasonable DSC of 0.2-0.8 and the diffeomorphic deformation improves the results to 0.3-0.9. Indicated by the positive DSC, both affine and diffeomorphic transformations sufficiently fill in the organs. The improved performance of diffeomorphic deformation is due to the fact that the affine transformation is linear with a limited degree of freedom. The diffeomorphic transformation is non-linear and more flexible, resulting in better anchor organ alignment.

This superior anchor organ framework results in a more accurate prediction of unsegmented organs. This effect can be further observed in that for non-anchor organs bordered by anchor organs in multiple directions, such as the trachea-bronchi, thymus, esophagus, and adrenals, the gain from the diffeomorphic deformation is higher. On the contrary, for non-anchor organs with limited constraints, such as the breast and larynx-pharynx, the DSCs were moderately improved from deformation.

**Registration leave-one-organ-out.** Fig. 5 shows the results of registration validation using the leave-one-organ-out approach. Quantitative results (Appendix Fig. A1) show a similar trend that both affine (DSC of 0.2-0.8 for both anchor and non-anchor organs) and diffeomorphic (DSC of 0.8-0.9 for anchor organs and DSC of 0.4-0.9 for non-anchor organs) transformation are sufficiently able to fill in organs, with superior performance from the diffeomorphic deformation. The affine transformation results are not substantially affected by whether the organ or its neighboring organs are anchors or left-out. This demonstrates that this non-linear transformation is mainly optimized for whole-body features rather than those from local context. The diffeomorphic deformation shows high transformation accuracy for most anchor organs. This considerable improvement is minimally affected regardless of which organ is left out, demonstrating the flexibility of this nonlinear method.

## B. Dosimetry validation

Fig. 6 shows representative simulated dose maps from a reference and the test set case from a segmentation validation experiment as well as a separate registration leave-one-phantom-out validation. For segmentation validation, the dose maps are similar between the reference and the test set phantoms, except for slight discrepancies between bone boundary regions, demonstrating general agreement in CT material segmentation between the *i*Phantom framework and the reference. For the registration validation, the dose maps of the phantom from the test set are similar to those of the reference, with the one generated with diffeomorphic deformation exhibiting better results than the affine method alone.

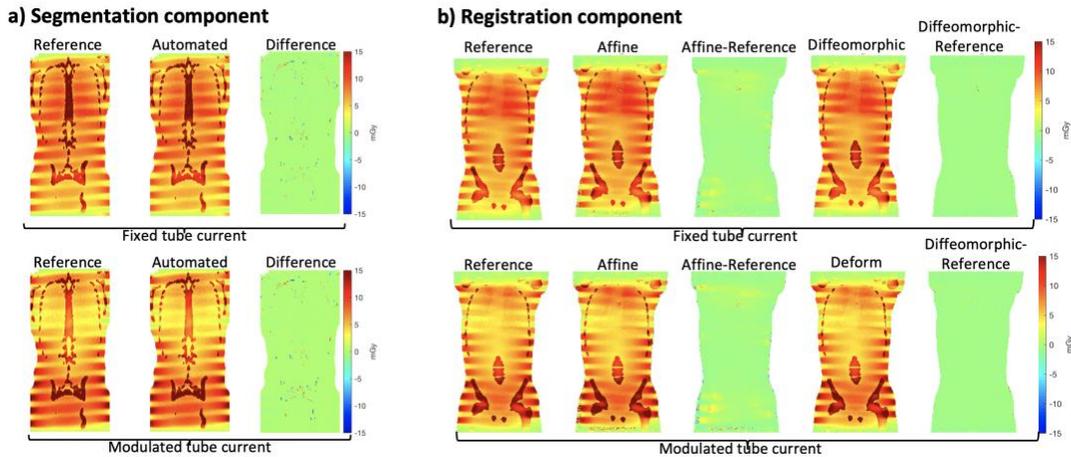

Fig. 6. Dose validation of the XCAT dataset. Dose maps of an example XCAT and the corresponding phantoms from the test set of the a) segmentation validation, and b) leave-one-phantom-out registration validation from simulated scans using fixed and modulated (bottom) tube current.

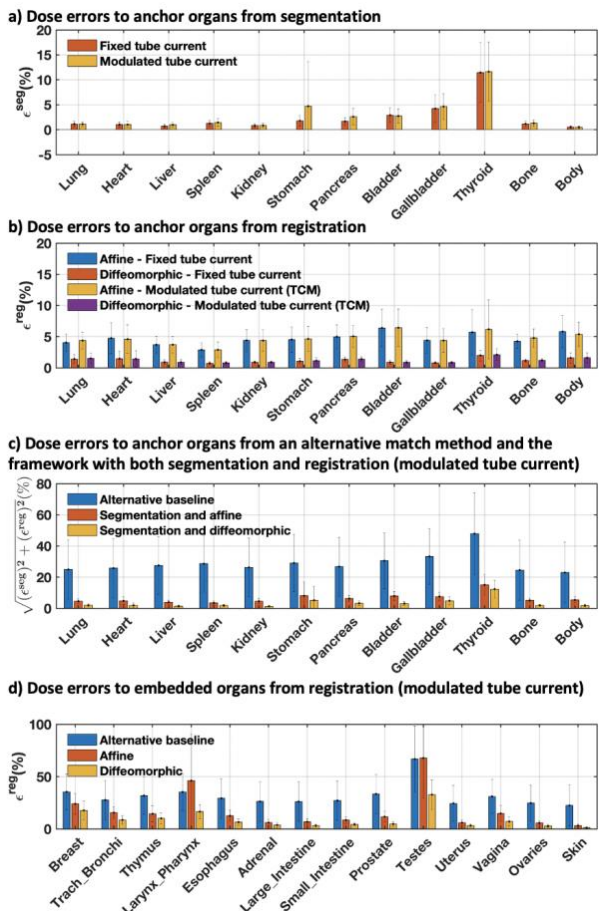

Fig. 7. Absolute relative error (%) of organ doses from phantoms from the test set and the reference averaged across the 50 XCAT for scans in leave-one-phantom-out validation. Organ dose errors to anchor organs result from a) segmentation of anchor organs, b) registration of non-anchor organs, and c) combining segmentation (a) and registration (b) errors or an alternative baseline method. d) Organ dose errors to embedded organs result from registration. The results on c) and d) are based on tube current modulated (TCM) scans, with the results from fixed tube current shown in the appendix.

Fig. 7 shows the mean absolute relative error between reference phantoms and their corresponding predicted anatomy from the *i*Phantom framework averaged across the 50 XCAT patients. For the segmentation validations (Fig. 7a), in general, anchor organs with DSC > 0.85 (lung, heart, liver, spleen, kidneys and bones) show a MAE of 0.5-1.5% for both fixed and modulated tube current scans. Organs challenging to segment (stomach, pancreas, bladder, gallbladder) show an average error of 1.5-4.5% and 2.5-5% for fixed and modulated scans, respectively. The MAE for thyroid is 12% for both fixed and modulated tube current. The dose errors from segmentation are generally small.

The filled-in organs introduce dose errors not only to the filled-in organs, but also to the anchor organs, resulting in anchor organs with a MAE of 2.5-6.5% by affine transformation, and 0.5-2.5% by diffeomorphic deformation for both fixed and modulated tube current (Fig. 7b). Both affine and diffeomorphic transformation show relatively small errors, with errors from diffeomorphic transformation being comparable to those from segmentation.

The overall errors to anchor organs combining segmentation (Fig. 7a) and registration (Fig. 7b) are 3-8% for affine transformation, and 1.5-5.5% for diffeomorphic deformation for both fixed and modulated tube current (Fig. 7c). Except for the thyroid, the MAE from the alternative baseline method (Section IV. E) are 7-14.5% for fixed tube current (Appendix Fig. A2a), and 23 − 33.5% for modulated tube current (Fig. 7c). The results demonstrate the sufficient and superior performance of the proposed approach for anchor organs.

Fig. 7d show dose errors for non-anchor organs. For fixed tube current (Appendix Fig. A2b), except for small organs or organs not fully constrained by anchors (breasts, larynx-pharynx, and testes), the affine method shows comparable results to the alternative baseline method (MAE: affine 3-18%, match 3-19%), with the MAE reduced to 1.5-7.0% with diffeomorphic deformation. Breasts, larynx-pharynx, and testes show relatively large errors with substantially superior results from diffeomorphic deformation (MAE: match 15-61%, affine 18-67%, deform 14-32%).

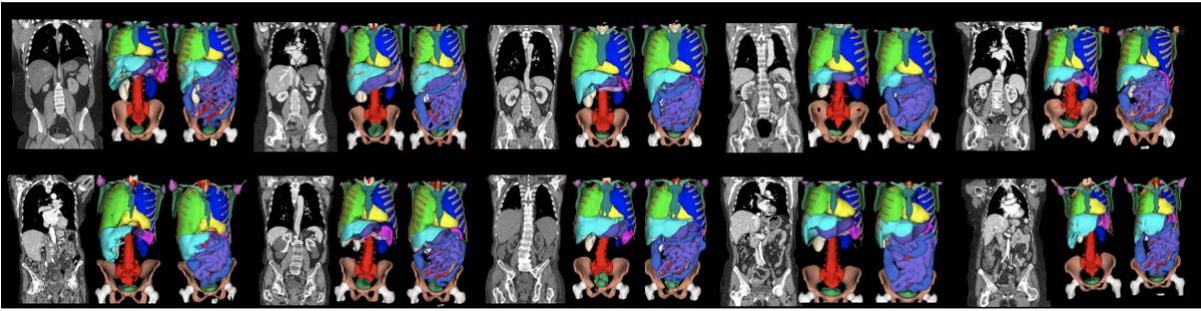

Fig. 8. Clinical validation. Rendering of the CT images, phantoms generated by automated segmentation and registration from the proposed framework.

For modulated tube current (Fig. 7d), compared to their corresponding MAE under fixed tube current, phantoms created by transformation show a slightly higher MAE while those by the alternative baseline method showed a substantially larger MAE (match 22-33.5%, affine 3-16%, deform 1-10%, except for the breasts, larynx-pharynx, and testes) for most organs. Results show that, except for limited types of organs, phantoms developed with the proposed transformations offer a high dose accuracy for the most embedded organs, substantially superior compared to those with the match-alone (alternative baseline) method, especially under modulated tube current.

For the alternative baseline method, the dose errors are drastically higher for scans with modulated tube current compared to the scans with fixed tube current. However, for phantoms created by registration (affine and diffeomorphic), the dose errors are only slightly higher for scans with modulated tube current compared to those with fixed tube current. For fixed tube current, by using phantoms with similar size, the organ dose errors are generally guaranteed within 20%, as widely demonstrated by the literature [19, 28]. However, the radiation field under modulated tube current is more heterogeneous, so it requires more precise organ location to obtain accurate dose values. Both affine and diffeomorphic transformation align the matched phantoms to the patients resulting in more accurate anatomical representation.

In the leave-one-organ-out validation, the MAE for filled-in organs, except for the thyroid, are 0-11% for affine, and 0-8% for diffeomorphic transformation for both modulated and fixed tube current (Appendix Fig. A3). The MAE is similar between scans with fixed and modulated tube current (<2%). Large filled-in organs and structures (lungs and bones) cause small dose errors on certain neighboring anchor organs (MAE: affine 0.5-4%, deform 0-2%). For example, when the lungs are left out and filled-in by diffeomorphic deformation, the lungs show a MAE of 1.4%, and the anchor organs show a MAE of 1.76% for the heart, 1.02% for the stomach, and 1.33% for the thyroid under tube current modulated scans. Other filled-in organs cause less substantial dose errors to the anchor organs with MAE <0.2% for both affine and deformation, for both fixed and modulated tube current. In general, the filled-in organs show sufficient dose values and a slight effect on the dose to anchor organs.

### C. Application to new clinical CT data

Fig. 8 shows renderings of the new phantoms created by applying the proposed framework to the clinical CT datasets. The renderings show that the segmentation and registration perform reasonably to build completed models. The quantitative results show similar trends as those from validation using the XCAT dataset (figures and tables in the appendix).

Compared to the XCAT dataset results, for organs easy to segment, the DSCs are $0.01 - 0.04$ less for lungs, heart, liver, spleen, kidneys, body, and $0.02 - 0.1$ less for bones. For organs with larger segmentation uncertainties (stomach, pancreas, bladder, gallbladder, and thyroid), the DSCs are -0.1 $- 0.12$ different compared to the XCAT validation (Appendix Table A6 and Fig. A4a). These differences are reasonable considering the annotation labels of the XCAT are created from different observers, and that the XCAT phantoms are largely based on normal patients, but the clinical patients were mostly abnormal. The differences in imaging techniques may also result in segmentation discrepancies.

The organ dose errors are comparable to the results from the XCAT dataset with simulations using tube current modulation, with a MEA of 0.5-1.5% for organs easier to segment, and a MEA of 1.5-5.5% for organs challenging to segment with detailed tables and figures shown in appendix (Appendix Fig. A4b and Table A7).

The DSC between phantoms from the test set and the reference in the leave-one-organ-out validation for the clinical dataset compared to the XCAT (Appendix Fig. A5a) is $0 - 0.2$ less on the average DSC (except for gallbladder, arms, and scapulars) for both affine and diffeomorphic transformation. The inferior results in the clinical dataset compared to the XCAT dataset may result from some large patients in the clinical dataset (effective diameter range: 24-44 cm for clinical patients versus 24-39 cm for the 50 XCAT patients). For both affine and diffeomorphic transformation, the anchor organ transformation calculation is unaffected by the type of organs left out, similar to the XCAT validation. For both affine and diffeomorphic transformation, the fill-in accuracy is not sensitive to whether the anchor organs are generated from the reference or predicted from the *i*Phantom framework. This demonstrates that the segmentation errors will not generally affect the fill-in accuracy compared to the ground truth.

For dosimetry (Appendix Fig. A5b), the MAE for filled-in organs are similar to those of the XCAT, except for the thyroid, with 0-11% for affine transformation, and 0-8% for diffeomorphic deformation with anchor organs from both the reference and the *i*Phantom prediction. The MAE for anchor organs, when obtained using the *i*Phantom predicted segmentation, is similar to the XCAT anchor organ overall error: 1.5-8% (except for gallbladder) for affine translation, and 1.5-5.0% (except for bladder) for diffeomorphic deformation.

## VI. DISCUSSION

In this study, we developed and validated a framework, *i*Phantom, to automatically generate computational human phantoms for individual patients, thus creating a digital "twin" for a patient based on his/her image data. We demonstrate that the framework provides a good tool for patient-specific organ dosimetry in CT. This specific application targets individualized or population-based image monitoring and protocol optimization. It shows the potential to efficiently generates large population of CHPs based on patient data.

The purpose of this study is to develop an overall framework architecture, while each module can be further optimized or customized. To improve the segmentation component, the number of training dataset, neural networks architecture, and algorithms can be enhanced. For the registration component, the template library, the registration algorithm, and the template matching criteria can also be updated based on the users' preference. Further, based on the application and the resources, the choice of anchor organs and all organs of interest can be further changed.

For patient-specific organ dosimetry, we combined the framework with a Monte Carlo simulation tool in a fully automated approach. State of the art studies providing similar ranges of organs usually approximate both anatomies (e.g., using match method) and radiation field. One study combined match-based anatomy modeling and a convolution-based radiation field modeling method, and it reported a dose error of about 20-35% [28].

The accuracy of the proposed framework is substantially superior with a MAE less than 10% for most organs even under the more challenging modulated tube current situation. It generally regarded that organ doses are sufficient with a 10% error limited by the Monte Caro simulation accuracy. Our results further suggest that for dosimetry applications, it is beneficial to apply a tiered approach: When computational resources are sparse, affine transformation offers reasonable accuracy; when computational resources are available, diffeomorphic deformation provides superior results; for the organs outside the field of view, one may use the alternative baseline method.

This study has several limitations. First, the dataset is relatively small. Second, the segmentation accuracy may be inferior compared to the state-of-the-art. The XCAT dataset is satisfactory for this task with a relatively large number of types of organs segmented and corresponding patient-specific phantoms. However, the dataset was developed ten years ago. Although we improved the quality of manual delineation for this study, due to the relatively low CT image quality and a large amount of organ annotation tasks (e.g. bones), the manual segmentation quality and CT image quality are not optimum. Third, the overall errors of the framework were approximated by combining separated errors from segmentation and deformation. These limitations are pathways for future improvements of the *i*Phantom methodology with present demonstrated quality and capability.

## VII. CONCLUSION

Computational human phantoms (CHPs) are essential for personalized clinical investigations and population-based simulation studies. However, their utility and generalization have been limited by current approaches in creating CHPs using manual segmentation. In this study, we proposed a novel framework, *i*Phantom, for automated and accurate creation of patient-specific CHPs from patient medical images. We showed that the framework precisely localized a wide range of organs, including low contrast organs, in CT images. Specifically, we presented an integrated framework built on fusing patient-specific automated learning-based segmentation with anatomical templates through template matching and diffeomorphic deformation. This framework was applied to patient-specific organ dosimetry, yielding a high accuracy (< 10% organ dose error) across radiosensitive organs. The components of the framework are modular and thus each can be further optimized for customized applications. This methodology may be useful for other applications, for example when dealing with hard-to-segment organs, lack of initial training data, and organ-based image quality evaluation.

## REFERENCES

[1] W. Kainz *et al.*, "Advances in Computational Human Phantoms and Their Applications in Biomedical Engineering—A Topical Review," *IEEE transactions on radiation and plasma medical sciences,* vol. 3, no. 1, pp. 1-23, 2018.

[2] X. G. Xu, "An exponential growth of computational phantom research in radiation protection, imaging, and radiotherapy: a review of the fifty-year history," *Physics in Medicine & Biology,* vol. 59, no. 18, p. R233, 2014.

[3] N. Petoussi-Henss, M. Zankl, U. Fill, and D. Regulla, "The GSF family of voxel phantoms," *Physics in Medicine & Biology,* vol. 47, no. 1, p. 89, 2001.

[4] C. Lee, J. L. Williams, C. Lee, and W. E. Bolch, "The UF series of tomographic computational phantoms of pediatric patients," *Medical physics,* vol. 32, no. 12, pp. 3537-3548, 2005.

[5] X. G. Xu, V. Taranenko, J. Zhang, and C. Shi, "A boundary-representation method for designing whole-body radiation dosimetry models: pregnant females at the ends of three gestational periods—RPI-P3,-P6 and-P9," *Physics in Medicine & Biology,* vol. 52, no. 23, p. 7023, 2007.

[6] A. Badano *et al.*, "In silico imaging clinical trials for regulatory evaluation: initial considerations for VICTRE, a demonstration study," in *Medical Imaging 2017: Physics of Medical Imaging,* 2017, vol. 10132: International Society for Optics and Photonics, p. 1013220.

[7] T. G. Schmidt, A. S. Wang, T. Coradi, B. Haas, and J. Star-Lack, "Accuracy of patient-specific organ dose estimates obtained using

an automated image segmentation algorithm," *Journal of Medical Imaging,* vol. 3, no. 4, p. 043502, 2016.

[8] W. P. Segars, B. M. Tsui, J. Cai, F.-F. Yin, G. S. Fung, and E. Samei, "Application of the 4-D XCAT phantoms in biomedical imaging and beyond," *IEEE transactions on medical imaging,* vol. 37, no. 3, pp. 680-692, 2017.

[9] J. Cho *et al.*, "Machine learning powered automatic organ classification for patient specific organ dose estimation," in *Proceedings of the Society for Imaging Informatics in Medicine Annual Meeting*, 2017.

[10] Z. Peng *et al.*, "A Method of Rapid Quantification of Patient‐Specific Organ Doses for CT Using Deep‐Learning based Multi‐Organ Segmentation and GPU‐accelerated Monte Carlo Dose Computing," *Medical Physics,* 2020.

[11] E. Gibson *et al.*, "Automatic multi-organ segmentation on abdominal CT with dense v-networks," *IEEE transactions on medical imaging,* vol. 37, no. 8, pp. 1822-1834, 2018.

[12] N. Pawlowski *et al.*, "Dltk: State of the art reference implementations for deep learning on medical images," *arXiv preprint arXiv:1711.06853,* 2017.

[13] W. Segars *et al.*, "Population of anatomically variable 4D XCAT adult phantoms for imaging research and optimization," *Medical physics,* vol. 40, no. 4, p. 043701, 2013.

[14] W. Segars *et al.*, "The development of a population of 4D pediatric XCAT phantoms for imaging research and optimization," *Medical physics,* vol. 42, no. 8, pp. 4719-4726, 2015.

[15] D. J. Tward *et al.*, "Patient specific dosimetry phantoms using multichannel LDDMM of the whole body," *International journal of biomedical imaging,* vol. 2011, 2011.

[16] W. Fu, W. P. Segars, E. Abadi, S. Sharma, A. J. Kapadia, and E. Samei, "From patient-informed to patient-specific organ dose estimation in clinical computed tomography," in *Medical Imaging 2018: Physics of Medical Imaging*, 2018, vol. 10573: International Society for Optics and Photonics, p. 1057315.

[17] Ö. Çiçek, A. Abdulkadir, S. S. Lienkamp, T. Brox, and O. Ronneberger, "3D U-Net: learning dense volumetric segmentation from sparse annotation," in *International conference on medical image computing and computer-assisted intervention*, 2016: Springer, pp. 424-432.

[18] S. A. Taghanaki *et al.*, "Combo loss: handling input and output imbalance in multi-organ segmentation," *Computerized Medical Imaging and Graphics,* vol. 75, pp. 24-33, 2019.

[19] X. Tian, X. Li, W. P. Segars, D. P. Frush, and E. Samei, "Prospective estimation of organ dose in CT under tube current modulation," *Medical physics,* vol. 42, no. 4, pp. 1575-1585, 2015. [Online]. Available: https://www.ncbi.nlm.nih.gov/pmc/articles/PMC4379759/pdf/MP HYA6-000042-001575_1.pdf.

[20] X. Li *et al.*, "Effects of protocol and obesity on dose conversion factors in adult body CT," *Medical physics,* vol. 39, no. 11, pp. 6550-6571, 2012.

[21] A. Sotiras, C. Davatzikos, and N. Paragios, "Deformable medical image registration: A survey," *IEEE transactions on medical imaging,* vol. 32, no. 7, pp. 1153-1190, 2013.

[22] B. B. Avants, C. L. Epstein, M. Grossman, and J. C. Gee, "Symmetric diffeomorphic image registration with cross-correlation: evaluating automated labeling of elderly and neurodegenerative brain," *Medical image analysis,* vol. 12, no. 1, pp. 26-41, 2008.

[23] B. B. Avants, N. J. Tustison, G. Song, P. A. Cook, A. Klein, and J. C. Gee, "A reproducible evaluation of ANTs similarity metric performance in brain image registration," *Neuroimage,* vol. 54, no. 3, pp. 2033-2044, 2011.

[24] J. Pluta, B. B. Avants, S. Glynn, S. Awate, J. C. Gee, and J. A. Detre, "Appearance and incomplete label matching for diffeomorphic template based hippocampus segmentation," *Hippocampus,* vol. 19, no. 6, pp. 565-71, Jun 2009, doi: 10.1002/hipo.20619.

[25] S. Sharma, A. Kapadia, W. Fu, E. Abadi, W. P. Segars, and E. Samei, "A real-time Monte Carlo tool for individualized dose estimations in clinical CT," *Physics in Medicine & Biology,* vol. 64, no. 21, p. 215020, 2019.

[26] A. Badal and A. Badano, "Accelerating Monte Carlo simulations of photon transport in a voxelized geometry using a massively parallel graphics processing unit," *Medical physics,* vol. 36, no. 11, pp. 4878-4880, 2009.

[27] X. Li, W. P. Segars, and E. Samei, "The impact on CT dose of the variability in tube current modulation technology: a theoretical investigation," *Physics in Medicine & Biology,* vol. 59, no. 16, p. 4525, 2014.

[28] X. Tian, W. P. Segars, R. L. Dixon, and E. Samei, "Convolution-based estimation of organ dose in tube current modulated CT," *Physics in Medicine & Biology,* vol. 61, no. 10, p. 3935, 2016.

# iPhantom: a framework for automated creation of individualized computational phantoms and its application to CT organ dosimetry – *Supplementary Materials*

Wanyi Fu, Shobhit Sharma, Ehsan Abadi, Alexandros Illiopoulos, Qi Wang,

Joseph Y. Lo, Xiaobai Sun, William P. Segars, Ehsan Samei

## I. SUPPLEMENTARY MATERIALS

### A. Anchor organ segmentation – 3D Unet architecture

The 3D Unet architecture consists of four encoder blocks followed by three decoder blocks. Each encoder block consists of two 3x3x3 convolution layers each followed by a leaky rectified linear unit (LeakyRelu) [1]. Except for the last encoder block, the last layer for each encoder block is 2x2x2 max pooling with a stride of 2, with the number of feature maps doubled before max pooling. Each decoder block consists of a 3x3x3 transpose convolution layer with a stride of 2 and a concatenating layer that combines feature maps from the output of the encoder block with the same resolution. The concatenated feature maps are followed by two 3x3x3 convolution layers each followed by a leaky rectified linear unit (LeakyRelu). The padding is used for all convolution

This work was supported in part by the Research Grant through the National Institutes of Health under Grant R01EB001838.

W. Fu and E. Abadi are with the Department of Electrical and Computer Engineering, and Carl E. Ravin Advanced Imaging Laboratories, Duke University, Durham, NC, 27705 USA (email: wanyi.fu@duke.edu; ehsan.abadi@duke.edu).

S. Sharma is with the Department of Physics and Carl E. Ravin Advanced Imaging Laboratories, Duke University, Durham, NC, 27705 USA (email: shobhit.sharma@duke.edu).

A. Illiopoulos and X. Sun are with the Department of Computer Science, Duke University, Durham, NC, 27708, USA (email: ailiop@cs.duke.edu; xiaobai@cs.duke.edu).

Q. Wang with the Department of Radiology, the Fourth Clinical Hospital of Hebei Medical University, Heibei, 050011, China (email: wq20@hotmail.com)

J. Y. Lo is with the Departments of Electrical and Computer Engineering, Biomedical Engineering, Medical Physics Graduate Program, and Carl E. Ravin Advanced Imaging Laboratories, Duke University, Durham, NC, 27705 USA (email: joseph.lo@duke.edu).

W. P. Segars is with the Departments of Biomedical Engineering, Medical Physics Graduate Program and Radiology and the Carl E. Ravin Advanced Imaging Laboratories, Duke University, Durham, NC, 27705 USA (e-mail: paul.segars@duke.edu).

E. Samei is with the Carl E. Ravin Advanced Imaging Laboratories, the Medical Physics Graduate Program, the Departments of Radiology, Electrical and Computer Engineering, Biomedical Engineering, and Physics, Duke University, Durham, NC, 27705 USA (email: ehsan.samei@duke.edu).

layers to preserve the size and information. The last layer is a Softmax function, with the number of the output set as the number of anchor organs types.

### B. Results of validation using the XCAT phantom dataset

In this section, we present supplemental figures and tables of iPhantom validation results using the XCAT dataset. The geometry validation results are shown in Tables A1-3 and Fig. A1 and the dosimetry validation results are shown in Tables A4-5 and Fig. A2-3. Tables A1 and A2 show the DSCs between the reference anatomy and the predicted using the iPhantom framework in the test set averaged across the 50 phantoms for the segmentation experiment ( $DSC^{seg}$ ) and registration leave-one-phantom out experiment ( $DSC^{reg}$ ) , respectively. Fig A.1. shows the average DSCs between the reference and the predicted using the iPhantom framework for each of the leave-one-organ out experiment ( $DSC^{reg}$ ). The corresponding results of Fig A.1. are summarized in Table A3, where the anchor organ sub-table shows the average $DSC^{reg}$ of anchor organs across all leave-one-organ-out experiments and the filled in organ sub-table shows the $DSCs^{reg}$ for filled-in organs in each leave-one-organ-out experiment .

Table A4 shows the mean absolute errors (MAE %) of anchor organ doses between reference phantom and the phantom predicted from the alternative baseline method ( $\varepsilon^{mat}$ ), the framework using segmentation component ( $\varepsilon^{seg}$ ), and the registration component ( $\varepsilon^{reg}$ ) in the leave-one-phantom out validation. Table A4 also shows the combined errors from the segmentation and registration component ( $\sqrt{(\varepsilon^{seg})^2 + (\varepsilon^{reg})^2}$ ). Table A5 shows the results for filled-in organs of the leave-one-phantom-out validation. The results in Tables A4-5 are also shown in Fig 8 in the main manuscript and Fig. A2 (fixed tube current). Figure A3 shows the dose errors from the leave-one-organ out validation experiment.

### C. Results of validation using the clinical dataset

The results of iPhantom validation using the clinical dataset are shown in Tables A6-7 and Fig A4-5. Table A6 and Fig A4a show the DSCs of segmentation between the predicted from the iPhantom framework and the reference. Table A7 and Fig. A4b shows the organ dose errors from the segmentation component. Fig. A5 shows the results of registration component in terms of DSCs (Fig. A5a) and MAE (Fig. A5b) between the predicted from the proposed framework and the reference, respectively. The results include using both reference and predicted segmentation as anchors.

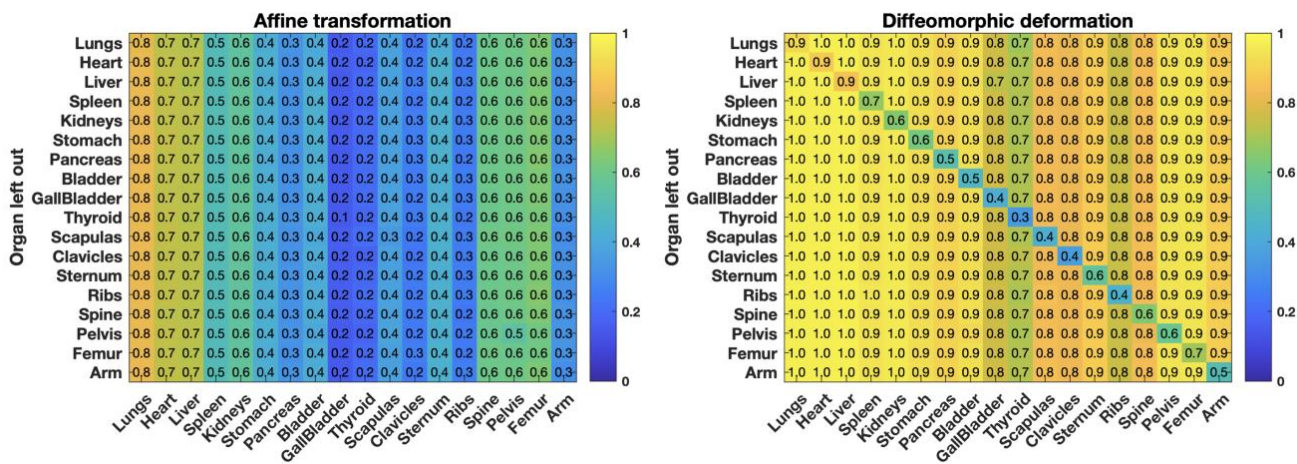

Fig. A1. Mean and standard deviation of dice similarity coefficients ($DSC^{reg}$) between the reference and the predicted anatomy from the framework in the test set (left: affine, right: diffeomorphic) across the 50 XCAT patients. Each row represents an experiment with an organ (specified by row name) left out and then filled in. The column name indicates the organ of the DSC value. Elements on the diagonal are values for filled-in organs and the rest for anchor organs. For affine transformation, the results are not substantially affected by which organs are left-out or whether the organs belong to filled-in or anchor organs. The diffeomorphic deformation sufficiently aligns the anchor organs. The diffeomorphic deformation accuracy for filled-in organs is mainly affected by the organ size and the extent of contact with the anchor organs.

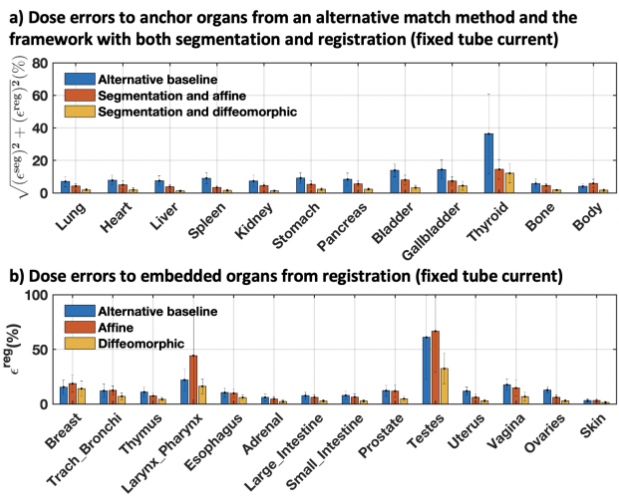

Fig. A2. Absolute relative error (%) of organ doses from phantoms based on the test set and the reference averaged across the 50 XCAT for scans in leave-one-phantom-out validation under fixed tube current. A) Organ dose errors to anchor organs result combining segmentation and registration and an alternative matching method. B) Organ dose errors to embedded organs result from registration.

## Leave one organ out results : dose errors from registration $\varepsilon^{reg}$

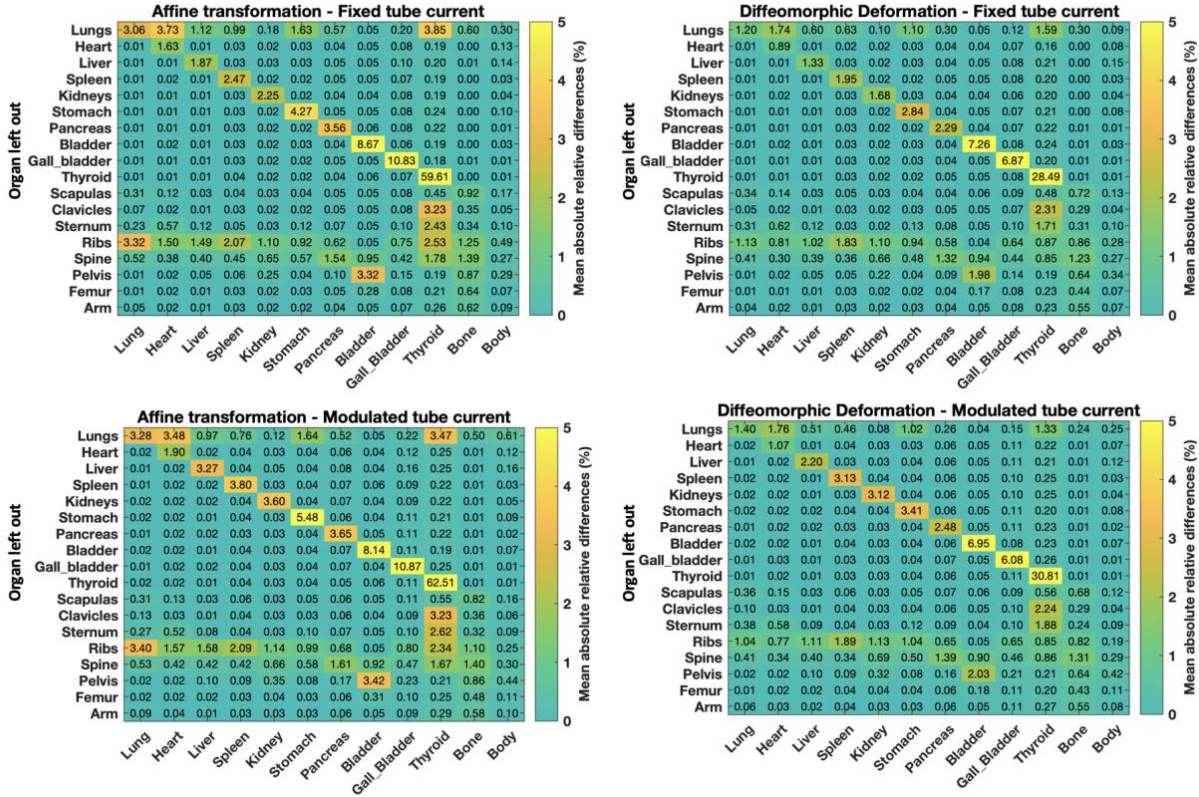

Fig. A3. Dosimetry validation – XCAT phantom results. Results from the leave-one-organ-out validation where each row represents an experiment with the specified organ out and each column represents the organ dose mean absolute errors (MAE %).

TABLE A1. RESULT OF IPHANTOM SEGMENTATION COMPONENT VALIDATION IN GEOMETRY. MEAN AND STANDARD DEVIATION OF DICE SIMILARITY COEFFICIENTS ($DSC^{seg}$) BETWEEN THE REFERENCE AND THE PREDICTED FROM THE PROPOSED FRAMEWORK IN THE TEST SET ACROSS THE 50 XCAT PATIENTS .

| Lung-R | Lung-L | Heart | Liver | Spleen | Kidney-R | Kidney-L | Stomach | Pancreas | Bladder | Gall-ladder |
|--------|--------|-------|-------|--------|----------|----------|---------|----------|---------|-------------|
| 0.99±0.01 | 0.98±0.01 | 0.93±0.04 | 0.95±0.01 | 0.91±0.05 | 0.90±0.08 | 0.90±0.10 | 0.77±0.16 | 0.57±0.15 | 0.81±0.11 | 0.61±0.20 |
| Spine | Ribs-R | Ribs-L | Clavicles | Sternum | Scapulars | Pelvis | Arm | Femur | Body | Thyroid |
| 0.94±0.02 | 0.88±0.04 | 0.88±0.04 | 0.90±0.04 | 0.86±0.07 | 0.92±0.05 | 0.95±0.02 | 0.90±0.10 | 0.95±0.02 | 0.98±0.01 | 0.59±0.15 |

TABLE A2. RESULT OF IPHANTOM REGISTRATION COMPONENT LEAVE-ONE-PHANTOM-OUT VALIDATION IN GEOMETRY. MEAN AND STANDARD DEVIATION OF DICE SIMILARITY COEFFICIENTS ($DSC^{reg}$) BETWEEN THE REFERENCE AND THE PREDICTED FROM THE FRAMEWORK IN THE TEST SET ACROSS THE 50 XCAT PATIENTS.

| | a. Anchor organs | | b. Filled-in organs | | |
|---|---|---|---|---|---|
| | Affine | Diffeomorphic | | Affine | Diffeomorphic |
| Lungs | 0.82±0.06 | 0.97±0.01 | Breast | 0.27±0.21 | 0.44±0.26 |
| Heart | 0.74±0.13 | 0.97±0.01 | Trachea bronchi | 0.28±0.18 | 0.57±0.16 |
| Liver | 0.73±0.11 | 0.97±0.01 | Thymus | 0.30±0.24 | 0.52±0.20 |
| Spleen | 0.50±0.20 | 0.94±0.03 | Larynx pharynx | 0.18±0.23 | 0.33±0.25 |
| Kidneys | 0.56±0.12 | 0.96±0.01 | Esophagus | 0.24±0.20 | 0.47±0.17 |
| Stomach | 0.43±0.17 | 0.93±0.04 | Adrenals | 0.14±0.19 | 0.45±0.18 |
| Pancreas | 0.32±0.20 | 0.90±0.05 | Large intestine | 0.39±0.17 | 0.50±0.13 |
| Bladder | 0.42±0.26 | 0.93±0.03 | Small intestine | 0.46±0.15 | 0.55±0.13 |
| Gallbladder | 0.16±0.25 | 0.76±0.24 | Prostate | 0.53±0.40 | 0.62±0.33 |
| Thyroid | 0.21±0.22 | 0.71±0.15 | Testes | 0.50±0.42 | 0.60±0.35 |
| Spine | 0.60±0.10 | 0.84±0.04 | Uterus | 0.75±0.37 | 0.85±0.22 |
| Ribs | 0.25±0.16 | 0.75±0.05 | Vagina | 0.77±0.34 | 0.82±0.25 |
| Clavicles | 0.25±0.24 | 0.82±0.12 | Ovaries | 0.74±0.38 | 0.77±0.34 |
| Sternum | 0.43±0.21 | 0.88±0.04 | Skin | 0.20±0.04 | 0.74±0.04 |
| Scapulars | 0.38±0.17 | 0.83±0.07 | | | |
| Pelvis | 0.61±0.15 | 0.94±0.01 | | | |
| Arm | 0.32±0.22 | 0.88±0.06 | | | |
| Femur | 0.63±0.20 | 0.95±0.02 | | | |

TABLE A3. RESULT OF IPHANTOM REGISTRATION COMPONENT LEAVE-ONE-ORGAN-OUT VALIDATION IN GEOMETRY. MEAN AND STANDARD DEVIATION OF DICE SIMILARITY COEFFICIENTS ($DSC^{reg}$) BETWEEN THE REFERENCE AND THE PREDICTED FROM THE FRAMEWORK IN THE TEST SET ACROSS THE 50 XCAT PATIENTS. FOR ANCHOR ORGANS, THE VALUE BELOW WAS THE MEAN AND STANDARD DEVIATION ACROSS ALL THE LEAVE-ONE-ORGAN OUT EXPERIMENTS IF THE ORGAN WAS USED AS ANCHOR ORGAN (NOTE THE STANDARD DEVIATION IS SMALL). FOR FILLED IN ORGANS, THE VALUE BELOW WAS DERIVED AS THE MEAN AND STANDARD DEVIATION WITHIN EACH LEAVE-ONE-ORGAN OUT EXPERIMENT WHEN THE ORGAN IS LEFT OUT.

| | Anchor organs | | Filled-in organs | |
|---|---|---|---|---|
| | Affine | Diffeomorphic | Affine | Diffeomorphic |
| Lungs | 0.82±0.00 | 0.97±0.00 | 0.8±0.07 | 0.91±0.03 |
| Heart | 0.74±0.00 | 0.97±0.00 | 0.74±0.13 | 0.88±0.06 |
| Liver | 0.73±0.00 | 0.97±0.00 | 0.72±0.12 | 0.86±0.05 |
| Spleen | 0.50±0.00 | 0.94±0.00 | 0.5±0.19 | 0.67±0.14 |
| Kidneys | 0.56±0.00 | 0.96±0.00 | 0.56±0.12 | 0.65±0.11 |
| Stomach | 0.43±0.00 | 0.93±0.00 | 0.43±0.17 | 0.62±0.13 |
| Pancreas | 0.32±0.00 | 0.90±0.00 | 0.32±0.21 | 0.52±0.14 |
| Bladder | 0.42±0.01 | 0.93±0.00 | 0.44±0.24 | 0.47±0.22 |
| Gall bladder | 0.16±0.00 | 0.76±0.01 | 0.15±0.24 | 0.42±0.28 |
| Thyroid | 0.20±0.01 | 0.71±0.01 | 0.19±0.22 | 0.35±0.18 |
| Spine | 0.60±0.00 | 0.84±0.00 | 0.59±0.11 | 0.64±0.08 |
| Ribs | 0.25±0.01 | 0.75±0.00 | 0.25±0.15 | 0.44±0.08 |
| Clavicles | 0.25±0.01 | 0.82±0.00 | 0.23±0.23 | 0.37±0.17 |
| Sternum | 0.44±0.01 | 0.88±0.00 | 0.43±0.19 | 0.57±0.11 |
| Scapulars | 0.37±0.00 | 0.83±0.00 | 0.35±0.17 | 0.43±0.14 |
| Pelvis | 0.61±0.00 | 0.94±0.00 | 0.55±0.17 | 0.59±0.12 |
| Arm | 0.32±0.01 | 0.86±0.09 | 0.3±0.22 | 0.5±0.24 |
| Femur | 0.64±0.01 | 0.95±0.00 | 0.61±0.21 | 0.69±0.14 |

TABLE. A4. RESULT OF IPHANTOM VALIDATION IN DOSIMETRY. MEAN ABSOLUTE ERROR (%) OF ANCHOR ORGAN DOSES BETWEEN THE REFERENCE AND THE PREDICTED FROM THE PROPOSED FRAMEWORK IN THE TEST SET WHEN THE PHANTOMS WERE CREATED USING THE ALTERNATIVE MATCHING METHOD ($\varepsilon_i^{mat}$), THE SEGMENTATION COMPONENT ($\varepsilon_i^{seg}$) AND REGISTRATION ($\varepsilon_i^{reg}$) COMPONENT. THE COMBINED ERROR $\sqrt{(\varepsilon^{seg})^2 + (\varepsilon^{reg})^2}$ WAS ALSO CALCULATED.

| | Fixed tube current | | | | | |
|---|---|---|---|---|---|---|
| Error types | Matched | Segment | Affine | Diffeomorphic | Segmentation and Affine | Segmentation and Diffeomorphic |
| | $\varepsilon^{mat}$ | $\varepsilon^{seg}$ | $\varepsilon^{reg}$ | $\varepsilon^{reg}$ | $\sqrt{(\varepsilon^{seg})^2 + (\varepsilon^{reg})^2}$ | $\sqrt{(\varepsilon^{seg})^2 + (\varepsilon^{reg})^2}$ |
| Lung | 7.04±6.12 | 1.14±1.04 | 4.04±2.72 | 1.41±1.45 | 4.30±2.77 | 1.91±1.68 |
| Heart | 7.78±5.97 | 1.03±1.14 | 4.75±4.97 | 1.47±2.38 | 4.94±5.01 | 1.91±2.56 |
| Liver | 7.53±6.10 | 0.70±0.71 | 3.71±2.65 | 0.90±0.78 | 3.84±2.64 | 1.21±0.96 |
| Spleen | 9.08±6.72 | 1.24±1.18 | 2.88±2.27 | 0.74±0.47 | 3.32±2.32 | 1.57±1.11 |
| Kidney | 7.31±7.01 | 0.82±0.83 | 4.41±3.51 | 0.89±0.43 | 4.58±3.49 | 1.30±0.79 |
| Stomach | 9.21±6.35 | 1.79±2.15 | 4.54±4.09 | 1.08±0.88 | 5.27±4.16 | 2.32±2.08 |
| Pancreas | 8.36±7.65 | 1.66±1.55 | 4.99±3.76 | 1.35±0.74 | 5.59±3.59 | 2.38±1.35 |
| Bladder | 13.93±8.08 | 2.91±2.96 | 6.40±6.03 | 0.88±0.59 | 8.03±5.73 | 3.24±2.82 |
| Gallbladder | 14.43±11.68 | 4.23±5.48 | 4.41±4.03 | 0.80±0.32 | 7.32±5.64 | 4.45±5.37 |
| Thyroid | 36.37±48.99 | 11.46±11.96 | 5.70±7.24 | 1.99±1.75 | 14.48±12.25 | 12.07±11.65 |
| Bone | 5.77±5.53 | 1.17±1.14 | 4.25±2.23 | 1.16±0.47 | 4.54±2.25 | 1.79±1.01 |
| Body | 4.08±2.96 | 0.52±0.49 | 5.82±5.22 | 1.59±1.62 | 5.87±5.21 | 1.74±1.62 |

| | Modulated tube current | | | | | |
|---|---|---|---|---|---|---|
| Error types | Matched | Segment | Affine | Diffeomorphic | Segmentation and Affine | Segmentation and Diffeomorphic |
| | $\varepsilon^{mat}$ | $\varepsilon^{seg}$ | $\varepsilon^{reg}$ | $\varepsilon^{reg}$ | $\sqrt{(\varepsilon^{seg})^2 + (\varepsilon^{reg})^2}$ | $\sqrt{(\varepsilon^{seg})^2 + (\varepsilon^{reg})^2}$ |
| Lung | 24.97±38.20 | 1.12±0.96 | 4.37±2.72 | 1.51±1.63 | 4.60±2.72 | 1.99±1.77 |
| Heart | 25.79±37.90 | 1.02±1.22 | 4.60±4.63 | 1.44±2.63 | 4.82±4.68 | 1.91±2.80 |
| Liver | 27.56±37.22 | 0.97±1.04 | 3.71±2.61 | 0.89±0.80 | 3.98±2.59 | 1.44±1.17 |
| Spleen | 28.66±36.67 | 1.45±1.52 | 2.89±2.42 | 0.79±0.46 | 3.55±2.44 | 1.80±1.42 |
| Kidney | 26.30±37.79 | 0.85±0.80 | 4.37±3.45 | 0.89±0.42 | 4.58±3.37 | 1.33±0.75 |
| Stomach | 29.09±36.66 | 4.70±17.86 | 4.64±4.10 | 1.15±0.91 | 8.08±17.71 | 5.20±17.78 |
| Pancreas | 26.77±37.51 | 2.60±3.25 | 5.04±3.50 | 1.40±0.74 | 6.31±3.88 | 3.25±3.04 |
| Bladder | 30.56±35.91 | 2.74±2.89 | 6.43±6.03 | 0.88±0.62 | 7.95±5.75 | 3.09±2.74 |
| Gallbladder | 33.33±35.70 | 4.64±5.17 | 4.37±3.81 | 0.85±0.38 | 7.58±5.11 | 4.87±5.04 |
| Thyroid | 47.91±52.46 | 11.64±11.83 | 6.18±9.43 | 2.11±1.91 | 15.09±13.26 | 12.35±11.44 |
| Bone | 24.54±38.47 | 1.30±1.31 | 4.77±2.95 | 1.21±0.50 | 5.09±2.99 | 1.93±1.17 |
| Body | 22.97±38.99 | 0.47±0.52 | 5.38±3.87 | 1.61±1.60 | 5.42±3.87 | 1.74±1.62 |

| | Fixed tube current | | | Modulated tube current | | |
|---|---|---|---|---|---|---|
| | Matched | Affine | Diffeomorphic | Matched | Affine | Diffeomorphic |
| | $\varepsilon^{mat}$ | $\varepsilon^{reg}$ | $\varepsilon^{reg}$ | $\varepsilon^{mat}$ | $\varepsilon^{reg}$ | $\varepsilon^{reg}$ |
| Breast | 15.51±12.49 | 18.67±15.85 | 14.08±14.22 | 35.39±34.58 | 23.95±20.02 | 17.71±17.64 |
| Trach Bronchi | 12.14±12.73 | 12.43±8.69 | 7.02±6.34 | 27.66±37.26 | 15.54±11.22 | 8.64±7.87 |
| Thymus | 10.99±8.89 | 7.38±6.00 | 4.30±3.49 | 31.89±35.98 | 14.42±15.27 | 10.09±10.23 |
| Larynx Pharynx | 22.05±20.14 | 44.33±80.85 | 16.19±13.58 | 35.31±35.25 | 46.14±89.51 | 16.49±13.55 |
| Esophagus | 10.46±8.65 | 9.82±7.58 | 5.97±4.63 | 29.30±37.58 | 12.62±10.24 | 6.47±6.45 |
| Adrenal | 6.14±6.59 | 4.63±4.35 | 2.26±2.18 | 26.19±37.87 | 6.04±6.19 | 3.67±4.97 |
| Large Intestine | 7.63±6.59 | 6.07±5.13 | 2.78±2.39 | 26.16±37.80 | 6.84±5.49 | 3.12±2.37 |
| Small Intestine | 7.89±7.11 | 6.41±5.21 | 2.74±2.15 | 27.11±37.38 | 8.50±6.46 | 4.41±3.01 |
| Prostate | 12.24±9.60 | 11.84±11.43 | 4.80±4.41 | 33.47±37.56 | 11.67±10.46 | 4.71±4.35 |
| Testes | 61.01±76.84 | 66.71±74.58 | 32.46±28.08 | 66.82±62.85 | 67.70±76.74 | 32.74±28.32 |
| Uterus | 11.87±7.60 | 6.13±5.95 | 3.03±2.52 | 24.36±34.30 | 5.97±5.55 | 3.24±2.59 |
| Vagina | 17.65±10.05 | 14.67±14.79 | 6.86±8.35 | 31.04±31.96 | 14.82±14.76 | 7.10±8.50 |
| Ovaries | 12.85±5.97 | 6.08±4.45 | 3.05±2.69 | 24.73±34.23 | 5.66±3.73 | 2.68±2.66 |
| Skin | 3.12±2.67 | 3.16±3.39 | 1.51±1.75 | 22.42±39.24 | 3.12±2.95 | 1.16±0.88 |

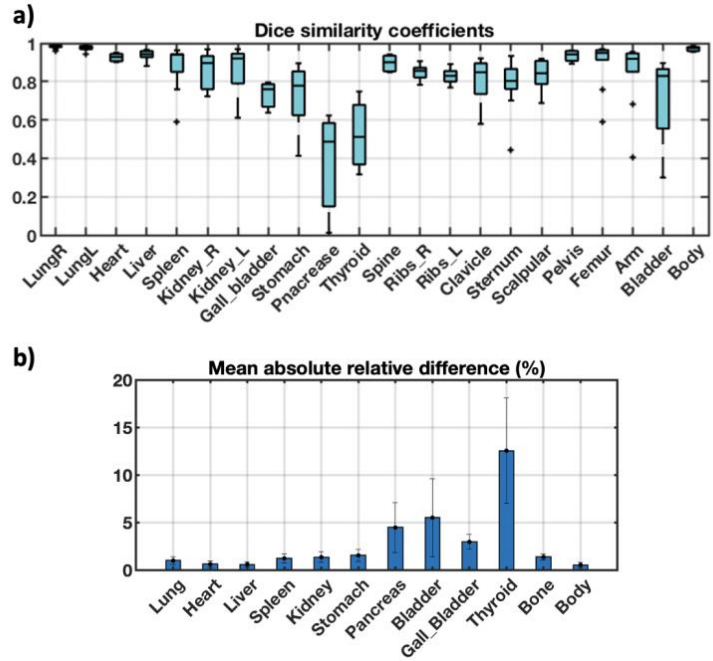

Fig. A4. Clinical validation of the segmentation components. a) Boxplots of DSC between the reference and the segmentation from the framework ($DSC_i^{seg}$) from the 10 datasets. b) Organ dose differences between phantoms based on the reference and the segmentation from the framework ($\varepsilon_i^{seg}$) averaged across the 10 datasets. Error bars represent ±1 standard deviation.

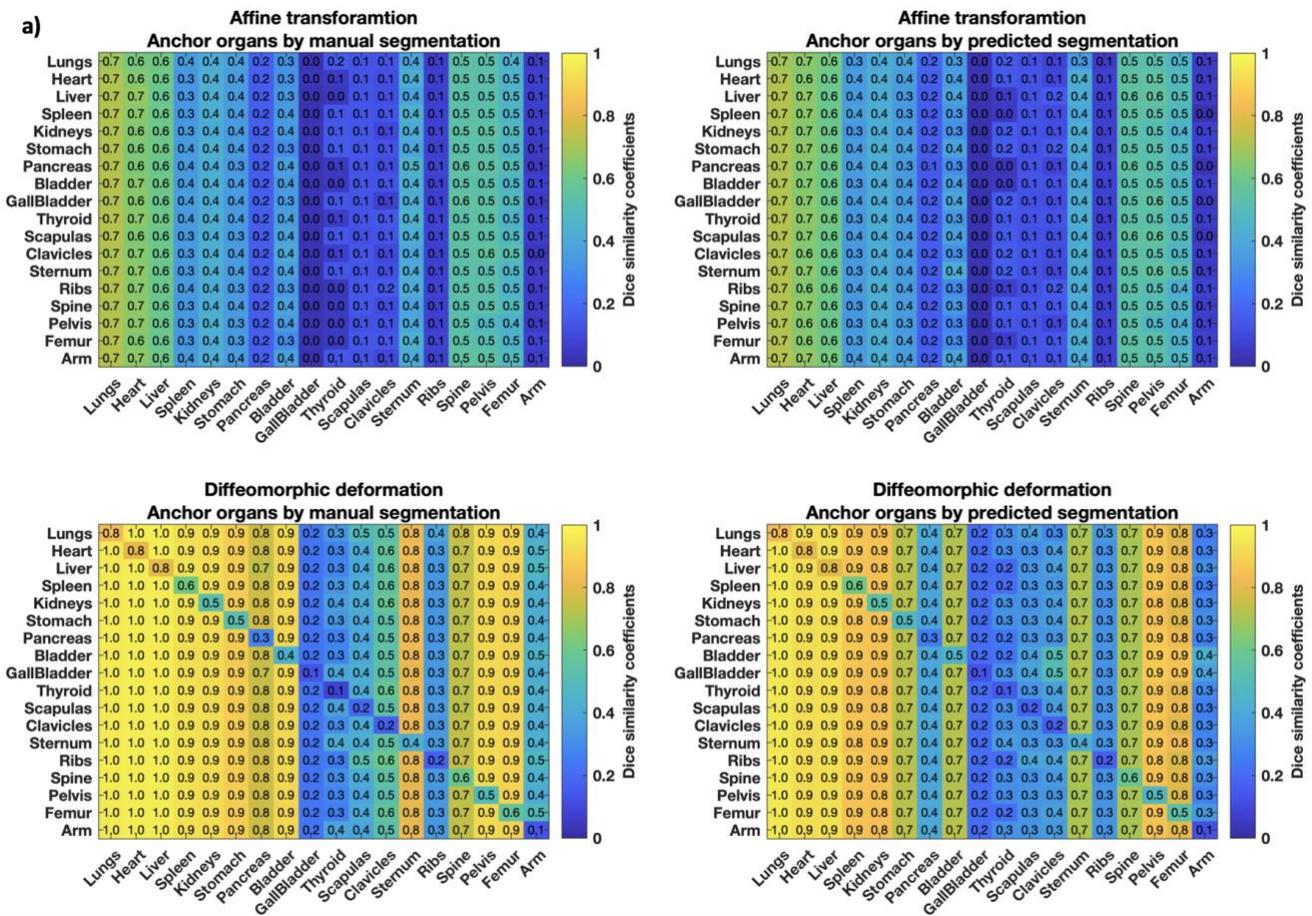

Fig. A5. a) Clinical validation of the registration components in terms of a) geometry $DSC_i^{reg}$. The results represent an average value from the 10 clinical datasets. The results show the DSC of reference phantoms and the phantoms created using affine transformation (top row) and diffeomorphic deformation (bottoms row) based on the reference (left columns) segmentation and automated segmentation (right columns) as anchors. Each row represents an experiment with the specified organ left out and then filled in and each column represents the value to the specified organs.

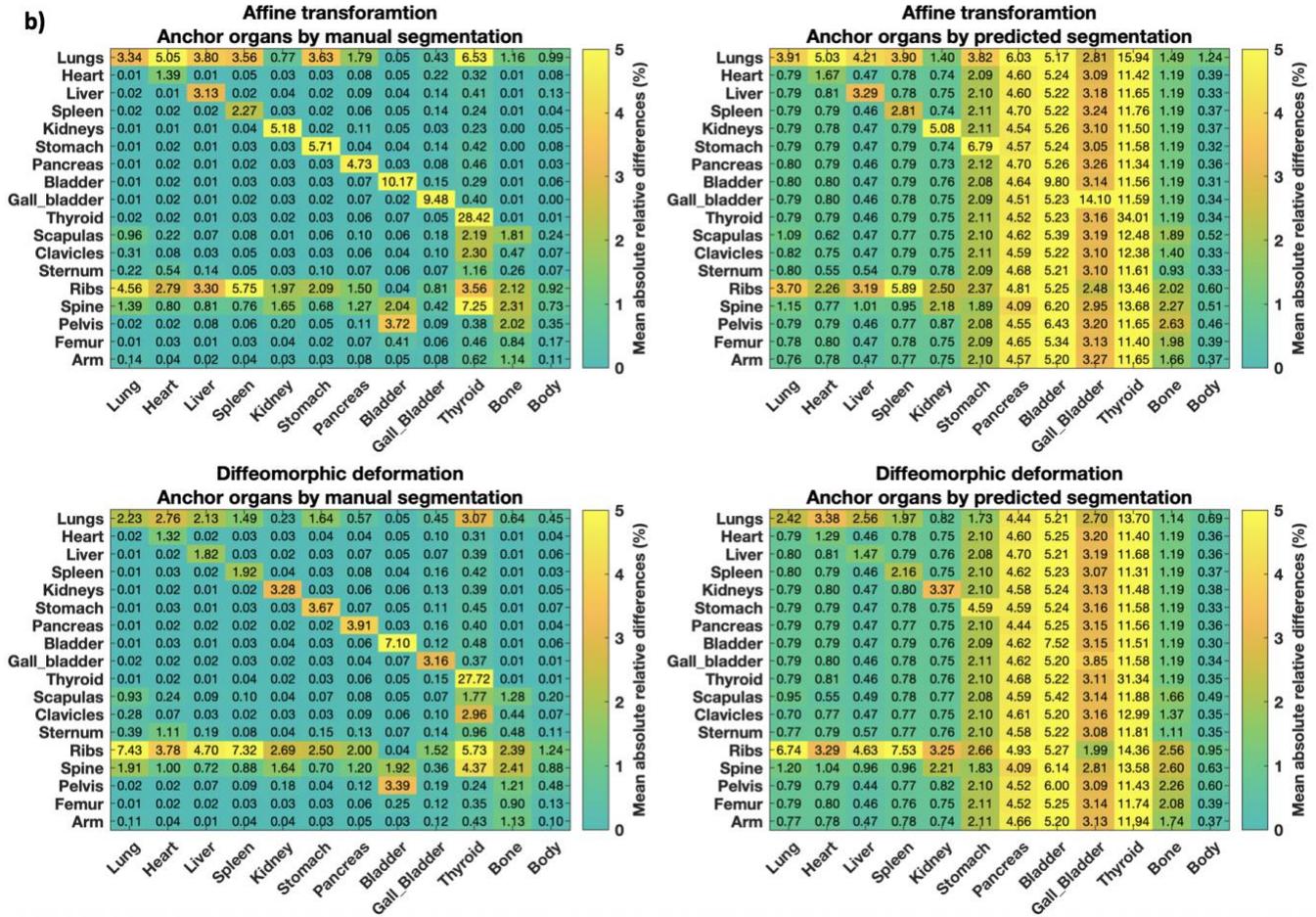

Fig. A5. b) Clinical validation of the registration components in terms of dosimetry $\varepsilon^{reg}$. The results represent an average value from the 10 clinical datasets. The results show the mean absolute dose errors (MAE %) of reference phantoms and the phantoms created using affine transformation (top row) and diffeomorphic deformation (bottoms row) based on the reference (left columns) segmentation and automated segmentation (right columns) as anchors. Each row represents an experiment with the specified organ left out and then filled in and each column represents the value to the specified organs.

TABLE A6. RESULT OF IPHANTOM SEGMENTATION COMPONENT VALIDATION IN GEOMETRY. MEAN AND STANDARD DEVIATION OF DICE SIMILARITY COEFFICIENTS ($DSC^{seg}$) BETWEEN THE REFERENCE AND THE PREDICTED FROM THE PROPOSED FRAMEWORK ACROSS THE TEN UNSEEN CLINICAL PATIENTS.

| Lung-R | Lung-L | Heart | Liver | Spleen | Kidney-R | Kidney-L | Stomach | Pancreas | Bladder | Gallbladder |
|---|---|---|---|---|---|---|---|---|---|---|
| 0.98±0.01 | 0.97±0.01 | 0.92±0.02 | 0.94±0.03 | 0.88±0.12 | 0.86±0.09 | 0.87±0.12 | 0.73±0.15 | 0.71±0.22 | 0.73±0.08 | |

| Spine | Ribs-R | Ribs-L | Clavicle | Sternum | Scapulars | Pelvis | Arm | Femur | Body | Thyroid |
|---|---|---|---|---|---|---|---|---|---|---|
| 0.90±0.04 | 0.85±0.04 | 0.83±0.04 | 0.81±0.11 | 0.78±0.14 | 0.83±0.08 | 0.93±0.03 | 0.85±0.18 | 0.90±0.12 | 0.97±0.01 | 0.52±0.16 |

TABLE A7. RESULT OF IPHANTOM SEGMENTATION COMPONENT VALIDATION IN DOSIMETRY. MEAN ABSOLUTE ERROR (%) OF ORGAN DOSES BETWEEN THE REFERENCE AND THE PREDICTED FROM THE PROPOSED FRAMEWORK SEGMENTATION MODULE ($\varepsilon_i^{seg}$)

| Lung | Heart | Liver | Spleen | Kidney | Stomach | Pancreas | Bladder | Gallbladder | Thyroid | Bone | Body |
|---|---|---|---|---|---|---|---|---|---|---|---|
| 0.98±0.79 | 0.61±0.70 | 0.59±0.49 | 1.22±0.99 | 1.34±1.14 | 1.53±1.27 | 4.46±5.29 | 5.50±8.21 | 2.95±1.52 | 12.54±11.08 | 1.36±0.71 | 0.53±0.46 |

REFERENCES

[1] Ö. Çiçek, A. Abdulkadir, S. S. Lienkamp, T. Brox, and O. Ronneberger, "3D U-Net: learning dense volumetric segmentation from sparse annotation," in International conference on medical image computing and computer-assisted intervention, 2016: Springer, pp. 424-432.